\documentclass[authoryear,preprint]{elsarticle}
\usepackage{graphicx}
\usepackage{epsfig}
\usepackage{amsmath}
\usepackage{tensor}
\usepackage{natbib}
\usepackage{xcolor}
\usepackage{amsfonts}
\usepackage[normalem]{ulem}

\usepackage{colortbl}
\usepackage{mathtools}
\usepackage{euscript}
\usepackage{mathbbol}

\newcommand{\be}{\begin{equation}}
\newcommand{\ee}{\end{equation}}
\newcommand{\bea}{\begin{eqnarray}}
\newcommand{\eea}{\end{eqnarray}}

\newcommand{\nd}{\noindent}

\begin{document}

\title{Counterintuitive properties of fixation probability and fixation time in population structures\\ with spatially periodic resource distribution}
\author[sharif]{Hossein Nemati}
\author[sharif]{Mohammad Reza Ejtehadi}
\author[uw]{Kamran Kaveh \corref{em}}
\cortext[em]{Corresponding author,\\
 E-mail address: kkavehma@gmail.com (K. Kaveh).}
\address[sharif]{Sharif University of Technology, Physics Department}
\address[uw]{University of Washington, Department of Applied Mathematics}

\date{\today}

\begin{abstract}
Resource are often not uniformly distributed within a population. Spatial variations of concentration of a resource, change the fitness of competing strategies locally. The notion of fitness varying with respect to both genotype and environment is important in modeling cancer initiation, microbial evolution and evolution of drug resistance. Environmental interactions can be asymmetric, that is, they affect the fitness of one type more than the other. The question is how local environmental variations in network population structures change the selection dynamics in a finite population setting. We consider one-dimensional lattice population structures with spatial fitness distributions with a periodic pattern. Heterogeneity is determined by standard deviation of fitnesses, $\sigma_{\rm A}, \sigma_{\rm B}$ and period $T$. The model covers biologically relevant limits of two-habitat subdivided populations and randomly-distributed resources in high- and low-periods. We numerically calculate fixation probability and fixation times for a constant population birth-death process as fitness heterogeneity and period vary. We identify levels of heterogeneity for which a previously deleterious mutant, in a uniform environment, becomes beneficial. In other regimes of the problem we observe unexpected behavior where the fixation probability of both types are larger than their neutral value at the same time: $\rho_{\rm A}, \rho_{\rm B} > 1/N$. This coincides with an exponential increase in time to fixation as a function of population size, which points to significant slow-down in selection process and the potential for coexistence between types in realistic time scales. We also discuss ‘fitness shift’ model where the fitness function of one type is identical to the other up to a constant spatial shift. This leads to significant increase (or decrease) in the fixation probability of the mutant depending the value of the shift. 
\end{abstract}

\maketitle

\section{Introduction}
Evolutionary dynamics is the study of how strategies arise and compete with each other. Success of a new mutant strategy depends on several factors. In unstructured populations, the fate of a mutant is primarily determined by its relative fitness compared to that of the resident population. Success of a mutant is measured in its ability to establish a finite colony (fixation) \citep{nowak2006evolutionary,broom2014game, ewens, traulsen2009stochastic, durrett2008probability, nagylaki1992introduction, lieberman2005evolutionary, allen2017evolutionary, ohtsuki2006simple}. 

Beside the inherent fitness of competing strategies, other factors such as population structure and environmental interactions can influence the outcome of the selection process. Population structure can represent spatial structure such as dispersal patterns in a population, a hierarchical phenotypic structure, or vicinity and neighborhood in social networks. Environmental interaction can represent the local changes of fitnesses of different strategies due to ecological factors such as interaction with local resources.

Evolutionary graphs, or network models are suggested as a powerful framework to study population structures \citep{lieberman2005evolutionary}. Population resides on nodes of a graph, $G$. Graph connectivities represent neighborhood topology and migration probability.  A category of graphs known as isothermal graphs is shown to have the same fixation probability as that of an unstructured population (complete graph) \citep{lieberman2005evolutionary, adlam2014universality}. (Also see \citep{maruyama1970fixation,maruyama1,maruyama2}.) Regular undirected graphs are examples of isothermal structures. Population structures that increase (decrease) the fixation probability of a beneficial mutant relative to that of unstructured populations are known as amplifier (suppressor) of selection respectively \citep{pavlogiannis2018construction,pavlogiannis2017amplification,adlam2015amplifiers}. Extensions to more general graph structures and/or update rules has been discussed in the literature \citep{komarova2006spatial,hindersin2014counterintuitive,hindersin2015most, broom2011evolutionary,broom2008analysis, monk2014martingales, houchmandzadeh2010alternative,houchmandzadeh2011fixation,allen2015molecular,hindersin2015most, kaveh2014duality,tkadlec2019population,pavlogiannis2017amplification}.

Fitness of each competing type or strategy is commonly assumed to be a function of its genotype or phenotype and decoupled from the population structure. Environmental interactions introduce an environment-induced component to the fitness. Variations in environmental conditions couples the fitness of types with the population structure and makes the prediction of outcome an evolutionary process tediously more difficult. Resource heterogeneity is a major example of local variations in environmental conditions. For example, in a population of E.coli bacteria, local concentration of nutrients (sugar) influences the fitness of E.coli. Higher concentration of nutrients lead to higher reproductive fitness. Competing E.coli strains metabolize different sugar types (glucose, lactose, etc).  Variation in a nutrient type affects the fitness of the strain with the corresponding metabolism \citep{thattai2003metabolic,lambert2014memory,kussell2005bacterial}. In the context of drug resistance, existence of a drug gradient increases the speed of evolution of resistant strains \citep{lambert2014bacteria,singh2010penetration,baym2016spatiotemporal}. This is discussed in the context of viral evolution in \citep{kepler1998drug,moreno2014imperfect} and cancer evolution in \citep{wu2013cell,fu2014spatial}.

Environment is modeled as a scalar function, $c_i$. $c_i$ is the resource concentration at location $i$. The distribution of resources can either be uniform spatially, or vary from location to location. Local interactions of each strategy with the local resources determines the fitness of that type at the given location. The change in fitness depends on the concentration of resource at that location. Environmental interactions can be asymmetric. This means that fitness of a mutant (type A) might not change the same amount as that of resident (type B) when encountered with the same concentration of resource. 

Discussing spatial variations in resources in structured populations requires a unified model that incorporates spatial structure and environmental interactions in one framework. In population genetics literature there has been much work in the context of subdivided meta-population models under diffusion approximation \citep{levins1963theory,gavrilets2002fixation,whitlock2005probability}. Environmental interaction in the context of evolutionary graph structures has been the focus of recent works in literature,\citep{kaveh2019environmental,kaveh2020moran,maciejewski2014environmental,giaimo2018invasion,mahdipour2017genotype,farhang2017effect,farhang2019environmental,masuda2010heterogeneous}
. Maciejewski et al, and Kaveh et al, suggested to represent a population structure with variable distribution of resource as a colored graph \citep{maciejewski2014environmental,kaveh2020moran}. Color of each node represents the concentration level of resource at that node. In general there can be as many colors as the nodes (locations) on the graph ($N$).

The question is how the interplay between population structure, represented by dispersal graph, $G$, and resource distribution, represented by a coloring map, $C$, affects the evolutionary advantage of a randomly placed mutant in a finite population setting. In general fitnesses are given by the vectors. ${\bf a} = (a_1,\cdots,a_N), {\bf b} = (b_1,\cdots,b_N)$, where $a_i (b_i)$ is the fitness of type A(B) at location $i$ respectively. 

In the current work we assume there are two resource levels, low-resource (poor) sites and high-resource (rich) sites (two-color scheme). The difference between fitnesses in poor and rich site denotes level of fitness heterogeneity in the system. Fitness heterogeneity level is proportional to the standard deviation of the fitness and  for type A (type B) and is denoted by $\sigma_{\rm A} (\sigma_{\rm B})$ respectively. In this simplified scheme we can write the fitness sets ${\bf a}, {\bf b}$ as,

\begin{align}
{\bf a} &= r_{\rm A} + \sigma_{\rm A}{\bf c}, \nonumber\\
{\bf b} &= r_{\rm B} + \sigma_{\rm B}{\bf c},
\label{fitness_rules}
\end{align}

\noindent where $r_{\rm A} (r_{\rm B})$ is the mean fitness of type A(type B) and ${\bf c} = (c_1,c_2,\cdots,c_N)$ is the resource function at every location. For two-concentration level scheme we assume $c_i = -1~(+1)$ for poor (rich)  resource sites respectively. We assume equal number of poor and rich sites. Notice that in Eq. \ref{fitness_rules}, $r_{\rm A}$ represents the fitness of type A plus the mean contribution from environment, i.e. uniform environment. For brevity, in the rest of the paper we simply call $r_{\rm A} (r_{\rm B}$) {\it inherent fitness} of type A (B) respectively.  If $r_{\rm A} > r_{\rm B}$ type A is {\it inherently beneficial} and if $r_{\rm A} < r_{\rm B}$ type A is {\it inherently deleterious}.

In finite populations, condition for selection of type A vs B is defined as when the fixation probability of a randomly placed type A in a background of type B's (resident), $\rho_{\rm A}$, is larger than the fixation probability of a type B in a background of type A's, $\rho_{\rm B}$,  

\begin{align}
\rho_{\rm A}({\bf a}, {\bf b}) > 1/N > \rho_{\rm B}({\bf a}, {\bf b}).
\label{CFS}
\end{align}

The question is how the changes in fitness standard deviations, $\sigma_{\rm A}, \sigma_{\rm B}$, changes the condition for selection. Especially, when the mean fitness of two types is kept the same, does increase in standard deviation makes one type advantageous relative to the other type? Furthermore, we can ask the same question when we keep mean fitnesses and fitness standard deviations the same but re-distribute the fitnesses spatially, i.e. use a different resource function ${\bf c}$. 

Another important question is that whether the condition for selection $\rho_{\rm A} > \rho_{\rm B}$ automatically implies $\rho_{\rm A} > 1/N$ and $\rho_{\rm B} < 1/N$. As we will see later, variable environment model breaks this condition and we can have $\rho_{\rm A} > 1/N$ and $\rho_{\rm B} > 1/N$ at the same time (or $\rho_{\rm A} < 1/N$ and $\rho_{\rm B} < 1/N$).

The answers to the above questions are complex and are unknown even in simple graph structures in the presence of environmental heterogeneities. To be able to answer the above questions, we further simplify the model by assuming linear graph structures and a periodic resource function, $c_i$. This implies that the fitnesses $a_i$ and $b_i$ are also periodic functions across the population. The periodic environment model is introduced so that we are able to characterize resource distribution, $c_i$, with just one parameter. This is the period of distribution, $T$, or the characteristic fitness length scale in the model, $c_i = c_{i+T}$. Thus the environmental heterogeneity is characterized with three characteristic scales: 1,2) standard deviation of fitnesses or heterogeneity levels ($\sigma_{\rm A,B}$) 3) Period of distribution $T$. This is the length scale that fitness variation occurs. 

The periodic fitness distribution has important biological and physical examples. When period is small, $T \sim 2$, rich and poor sites are well-mixed. This can be used as an approximation for fine-grained, random heterogeneities. When the system is divided into two geographically distinct environments, maximal period, $T=N$ is a good approximation. This is representative of two-island subdivided population models while unlike past works in literature (see for example \citep{gavrilets2002fixation}). However in our network model, the underlying population structure inside each island is taken into account - a linear graph is this case.

We discuss how condition for selection and fixation probability of a beneficial mutant ($r_{\rm A} > r_{\rm B}$)  (or neutral
$r_{\rm A} = r_{\rm B}$) changes with heterogeneity and period. For the condition for selection we consider a more general case that the heterogeneity, or inequality level, is different for each type ($\sigma_{\rm A}$ for type A vs $\sigma_{\rm B}$ for type B).  We determine for what range of values $\sigma_{\rm A}$ and $\sigma_{\rm B}$ a neutral mutant (neutral in uniform environment i.e. $r_{\rm A} = r_{\rm B}$) becomes advantageous or deleterious. We discuss similar conditions when $r_{\rm A} > r_{\rm B}$ (inherently advantageous mutant) or $r_{\rm A} < r_{\rm B}$ (inherently deleterious mutant).  We observe that there exist some spatial distributions in which almost any weak environmental interaction increases the fixation probability away from the uniform model with the exception of symmetric background fitness interaction.
 
Furthermore, we observe that for certain range of heterogeneities, the condition $\rho_{\rm A} > 1/N > \rho_{\rm B}$ is violated. We can now have a beneficial mutant, $\rho_{\rm A} > \rho_{\rm B}$, while fixation probabilities of both types are larger than the neutral case, $1/N$. This behavior is dominant when $\sigma_{\rm A} \approx - \sigma_{\rm B}$.   

Similarly, we discuss the fixation probability as the heterogeneity levels and period of fitnesses are 
varied. We consider two cases: 1) symmetric interactions, $\sigma = \sigma_{\rm A} = \sigma_{\rm B}$ (background fitness) and 2) fully asymmetric interactions, $\sigma = \sigma_{\rm A} = -\sigma_{\rm B}$ (opposite fitness). In both cases, for an inherently advantageous mutant in the most of periods the fixation probability is increased as the heterogeneity level $\sigma$ is increased. However the fixation probability of mutants is not ordered according to period. In other words, as period of fitness distribution is increased the fixation probability can decrease or increase. 

More interestingly, for the second scenario, while the fixation probability is increased the average time to fixation is increased by several orders of magnitude. This implies that while fixation probability of mutants are much higher than the residents (strong selection limit) the increased in time to fixation effectively put the selection process into a halt. This is a rare example of coexistence or polymorphism in finite population models. 

Finally, we investigate the scenarios where fitness distribution of both types are identical up to a phase-shift. We call this a fitness phase shift. This, for example, can be due to distribution of two different nutrients, periodically distributed, where each nutrient affects one type. We observe that in a neutral case, introducing a phase shift 
leads to significant change in fixation probability away from $1/N$.
 	
	
\section{Model} 

We consider two competing types or strategies, A (mutant) and B (resident or Wild Type). Both types reside on a cycle graph, $G$,  that represents the spatial structure. Each node of the graph denotes a location. Each location, $i$, is occupied by one strategy. We denote the occupancy by $n_i (n_i = 0,1)$. If $n_i = 1$ node $i$ is occupied by a mutant (strategy A) and if $n_i = 0$ there is a resident reside on that location. Total number of nodes is $N$. Edges of the graph represents the neighborhood. Connectivity is manifested in the form of migration matrix $m_{ij}$. If $m_{ij} \neq 0$ the two nodes $i$ and $j$ are connected ($i= j \pm1$). Furthermore, value of $m_{ij}$ represents the probability that an offspring of an occupant of the node $i$ migrates to a neighboring node $j$. For an undirected cycle $m_{ij} = (1/2)(\delta_{i,j+1} + \delta_{i,j-1})$,where $\delta_{ij}$ is Kronecker $\delta$-function.

\begin{figure}
\begin{center}
\includegraphics[width=0.8\textwidth]{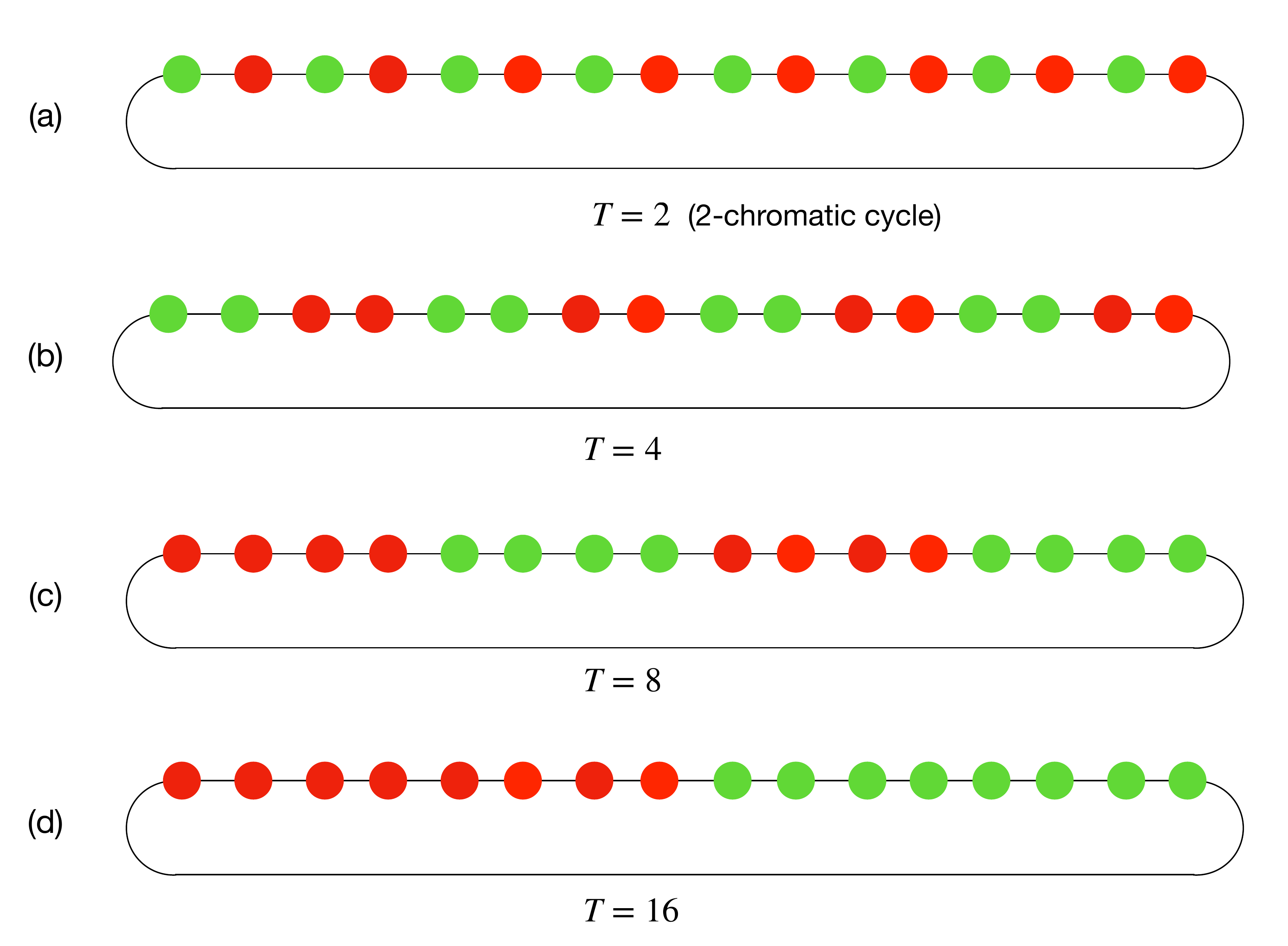}
\caption{\textbf{Periodic fitness distribution on a cycle.} Various configurations of rich- (green) and poor-resource sites (red) on a cycle of size $N=16$. Period of distribution, $T$, is varied between different configurations. a) $T=2$. b) $T=4$ c) $T=8$, d) $T=N=16$. At green (red) nodes the fitnesses are $a_{\rm green} = r_{\rm A} + \sigma_{\rm A}, b_{\rm green} = r_{\rm B} + \sigma_{\rm B} (a_{\rm red} = r_{\rm A} - \sigma_{\rm A}, b_{\rm red} = r_{\rm B} - \sigma_{\rm B})$ (respectively).}
\label{cycle}
\end{center}
\end{figure}
	
As discussed in the previous section, fitness of either types A or B, depends on both their strategy and the resource concentration level at their corresponding location. We use $r_{\rm A} (r_{\rm B})$ for inherent fitness of type A (type B) respectively, in the absence of environmental variations. (The uniform increase/decrease in the fitness of strategies is absorbed into $r_{\rm A,B}$). The resource function $c_i$ describes the concentration level of resource (nutrients or drug) at each location. For poor locations $c_i = -1$ and for rich locations $c_i =1$. This leads to the fitness map , Eq. \ref{fitness_rules}, i.e. $a_{i} = r_{\rm A} + \sigma_{\rm A}c_i,b_{i} = r_{\rm B} + \sigma_{\rm B}c_i$. Notice that we assumed equal number of rich and poor sites. Thus, the men fitnesses for A and B end up being $r_{\rm A}, r_{\rm B}$.

If the resource (at a given location) interacts with either types symmetrically, i.e. the fitness gain or loss due to the local resource is the same for each type, we have additive {\it background fitness} heterogeneity \citep{hauser2014heterogeneity}. In this case $\sigma_{\rm A}=\sigma_{\rm B} = \sigma$. The local resources affect the fitness of both types symmetrically,

\begin{align}
a_{i} &= r_{\rm A} + \sigma c_i ~~~~~ \textrm{(completely symmetric environmental interaction)}\nonumber\\
b_{i} &=r_{\rm B} + \sigma c_i \nonumber\\
\label{scenario1}
\end{align}

Similarly, we can consider a completely asymmetric interaction scenario, where $\sigma_{\rm A} = -\sigma_{\rm B} = \sigma$. In this case, a given resource concentration level at locatoin $i$, $c_i$, have opposite effect on the two strategies. At a (rich) location that fitness of type A is increased by $\sigma$ the fitness of B is indeed decreased by $-\sigma$ at the same site.  Fitness map in this asymmetric case is written as,  

\begin{align}
a_{i} &= r_{\rm A} + \sigma c_i~~~~~\textrm{(completely asymmetric environmental interaction)}\nonumber\\
b_{i} &= r_{\rm B} - \sigma c_i \nonumber\\
\label{scenario4}
 \end{align}

In both of the above scenarios, the resource function $c_i$ is periodic with period $T$, i.e. $c_{i+T} = c_i$. A schematic of the a graph of size 16 at different resource distributions $T=2,4,8,16$ is depicted in figure \ref{cycle}. Increase in period increases the segregation of rich and poor sites, i.e. at low-$T$ most rich sites are neighbor to poor sites while at large-$T$ most rich sites are neighbor to other fellow rich locations.  

\begin{figure}
\centering
\includegraphics[width=0.7\textwidth]{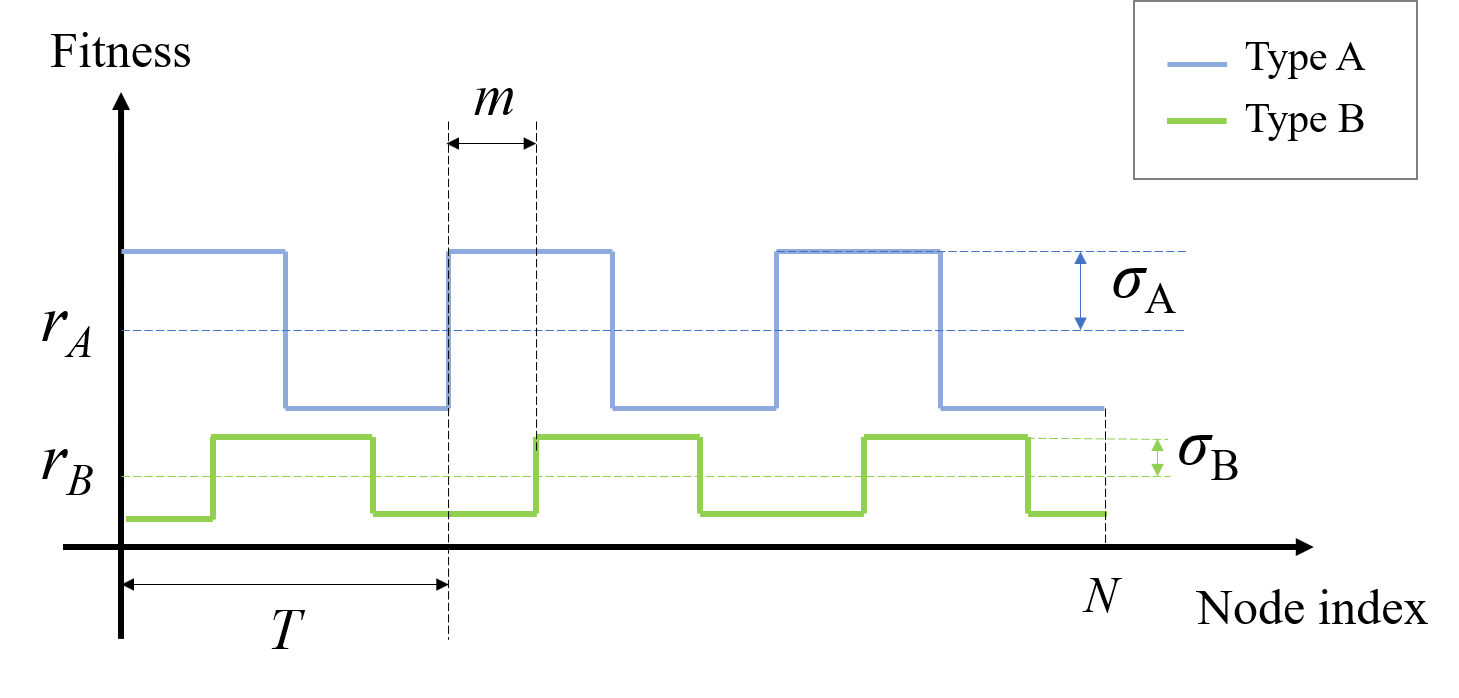}
\caption{\textbf{Periodic fitness patterns.} Each type has a square pattern of fitness with different means and amplitudes but the same period of $T$. There can also be a phase shift between the fitness pattern of the two types. }
\label{phase_diff_pattern}
\end{figure}

We can also consider fitness maps where $a_i$ and $b_i$ have identical spatial patterns up to a constant shift between the two patterns. Fitness shift could have relevant biological meaning, e.g. when each type feeds off from a separate nutrient or local resource. In the case of E.coli, we can have two competing strains which each metabolize a different sugar (say Glucose vs Lactose). (See \cite{lambert2014memory} and references therein.) The distribution of each of these resources can be spatially independent. It is curious to consider identical spatial distributions for each sugar type while one is merely different from the other by a constant shift in spatial coordinates. The reason we are interested to consider fitness distributions with a phase shift between $a_i$ and $b_i$ is that one can have identical mean and standard deviation (or higher moments) of fitness sets ${\bf a}$ and ${\bf b}$. The shift implies that $a_i = b_{i + m}$, where $m={\delta \over 2 \pi} T$ and $0 \geq \delta \geq 2\pi$ is the phase difference between $a_i$ and $b_i$. The evolutionary dynamics might not end up neutral since there is a non-trivial redistribution of fitnesses via fitness pattern shift $m$. The general fitness functions, $a_i$ and $b_i$, as a function of locations $i$ are plotted in figure \ref{phase_diff_pattern}. 
		
Except Section 3.3 which we focused on the effect of phase shift, we are mainly going to study the case of $m=0$ (in phase), which is equivalent to Eq. \ref{fitness_rules}. This means that there are some green and red nodes periodically in the cyclic graph, and each type gets an amount of excess fitness on the green nodes and lose the same amount of fitness on the red ones. 
		
\subsection{Numerical method; transition matrix}
For the dynamics we use a Moran birth-death updating. At every time step one individual chosen randomly and proportional to its fitness, $a_i$ or $b_i$, to reproduce. The offspring randomly replaces a neighboring individual. The birth-death update rules define a finite Markov chain with two absorbing states: fixation and extinction of the mutant; all the other states in the model are transient states. There is a standard method called \textit{transition matrix}\citep{grinstead} to solve such a Markov chain problem (see also \citep{hindersin2014counterintuitive}). 
	
Suppose that there are $t$ transient states and $a$ absorbing ones and totally $s=t+a$ states for the Markov chain. One can make up a stochastic matrix $\bold{T}_{s \times s}$ which contains the transition probabilities between any two states, i.e. $T_{ij}$ is the transition probability from state $i$ to state $j$ (do not confuse the indexing of the states with the indexing of the nodes). If we label the states in such an order that the first $t$ labels refer to the transient states and the last $a$ ones refer to the absorbing states, the shape of $\bold{T}_{s \times s}$ is like this: 
	
\begin{equation}
\bold{T}_{s \times s}=
\begin{pmatrix}
\bold{Q} & \bold{R} \\
\bold{0} & \bold{I}
\end{pmatrix}
\end{equation}
	
\nd in which, blocks $\bold{Q}_{t \times t}$ and $\bold{R}_{t \times a }$ contain the transition probabilities between the transient states and from the transient to the  absorbing states, respectively. The left bottom block, $\bold{0}_{a \times t}$ is a zero matrix, as there is no probability of transition from the absorbing states to the transient ones. The right bottom block $\bold{I}_{a \times a}$ is the identity matrix, because if the dynamics reaches an absorbing state, nothing will change anymore. This form of the transition matrix is known as \textit{Canonical form}. The procedure of obtaining $\bold{Q}$ and $\bold{R}$ matrices elements for a cycle graph is presented in \ref{Q_R_elemants}.
	
For this transition matrix, the fundamental matrix of the Markov chain is defined as:
	
\begin{equation}
\bold{F}=\sum_{n=0}^{\infty}{\bold{Q}^n}=(\bold{I-Q})^{-1},
\end{equation}
(see \cite{grinstead}).

The elements of this matrix have a special meaning. The element $F_{ij}$ is the average sojourn time in the state $j$ before the absorption, provided that the dynamics started from state $i$. The indices $i$ and $j$  refer to two transient states.
	
Having the matrices $\bold{F}$ and $\bold{R}$, we can have the matrix of absorption probabilities, which we call $\boldsymbol{\phi}$ and its size is ${t \times a}$:
	
\begin{equation}
\boldsymbol{\phi}=\bold{F}\bold{R}
\label{FR}
\end{equation}
	
\nd in which, the element $\phi_{ij}$ represents the probability of ultimate absorption to the $j$th absorbing state ($1\leq j \leq a$), provided that the dynamics is started from the $i$th transient state($1\leq i \leq t$). As in the present problem there is just two absorbing states, we can calculate the fixation and extinction probabilities of the mutant, starting from any initial state.
	
	As the elements of $\bold{F}$ represent the mean sojourn times, we can also calculate the mean time to absorption, starting from any $i$th transient state, which we call $\tau_i$:
	
\begin{equation}
\tau_i=\sum_{j=1}^{t}{F_{ij}}.
\end{equation}
	
The conditional absorption time which means that the mean time to reach a specific absorbing state, say the state $k$, starting from the $i$th transient state is \citep[Chapter~2]{ewens} \citep{altrock2012}:
\begin{equation}
\tau_{ik}=\sum_{j=1}^{t}{{\phi_{jk} \over \phi_{ik}}F_{ij}}.
\end{equation}
		
We labeled the states in such an order that the first $N$ labels refer to the configurations with a single mutant. In other words, by state $i$ ($1 \leq i \leq N$), we mean the specific configuration in which there is a mutant in the node $i$ and all the other nodes of the graph are occupied by the residents. In the canonical form we use here, the two absorbing states, totally resident and totally mutant configurations, among all the states are labeled as $s-1$ and $s$, respectively. But, $a=2$ and among the absorbing states they are labeled as the 1st and 2nd absorbing states. Thus, the fixation probability and average fixation time starting from a single mutant at the node $i$ are shown as $\phi_{i2}$ and $\tau_{i2}$. By definition, the fixation probability and fixation time are respectively:
\begin{equation}
\rho={1 \over N} \sum_{i=1}^{N}{\phi_{i2}}
\label{rho}
\end{equation}
and
\begin{equation}
\tau={1 \over N} \sum_{i=1}^{N}{\tau_{i2}}.
\label{tau}
\end{equation}
		
\section{Results}
	
\subsection{Evolutionary advantage and the fixation probability in a periodic environment}
We write the fixation probability in terms of fitness distribution parameters: $\rho_{\rm A} = \rho_{\rm A}(r_{\rm A}, r_{\rm B},\sigma_{\rm A}, \sigma_{\rm B}, T)$. The condition for selection of type A against B is defined as $\rho_{\rm A} > \rho_{\rm B}$. In a uniform environment, this is simply $r_{\rm A}>r_{\rm B}$. Condition for selection in a variable environments is exactly known in two limits; i) Complete graph in large-$N$ limit with any fitness distribution (any coloring) and ii) Properly two-colored regular graphs (two-chromatic graphs) at any $N$. For the case (i) the arithmetic mean fitness dictates the conditions for selection \citep{kaveh2019environmental}. That is, $\sum_{i} a_{i} > \sum_{i} b_{i}$. Conversely, for the two-chromatic graphs, case (ii), the geometric mean of the fitnesses of two types determines the condition for selection, i.e. $\prod_{i}a_{i} > \prod_{i} b_{i}$ \citep{kaveh2020moran}. Substituting $a_i$ and $b_i$ from Eq. \ref{fitness_rules}, and the fact the that $c_i=+1$ for half of nodes and $c_i = -1$ for the other half, we have, 

\begin{subequations}		
\begin{align}
&r_{\rm A} > r_{\rm B} ~~~~~~~~~~~~~~~~~\text{(Any-coloring complete graph)} \\
& r^{2}_{\rm A} - \sigma^2_{\rm A} > r^{2}_{\rm B} - \sigma^2_{\rm B} ~~~ \text{(Two-chromatic regular graphs)}
\label{condition_1}
\end{align}	
\end{subequations}

Eq. \ref{condition_1}, corresponds to the case of cycle with fitness period of $T=2$. In this case, if one type is inherently beneficial compared to the other type, i.e., $r_A \ne r_B$, (say $r_A=1.1$ and $r_B=1$) the curve of $\rho_{\rm A}=\rho_{\rm B}$ is a hyperbola in $\sigma_{\rm A}-\sigma_{\rm B}$ plane, Eq. \ref{condition_1}. For an inherently neutral mutant $(r_{\rm A} = r_{\rm B})$ this curve is composed of two diagonal lines $\sigma_{\rm A} = \pm \sigma_{\rm B})$. The neutrality curve is symmetric relative to the sign of $\sigma_{\rm A}$ and $\sigma_{\rm B}$. Notice that due to symmetry, the condition for selection of a deleterious mutant with say $r_{\rm A} =1, r_{\rm B}=1.1$ can be obtained with swapping the indices $A \leftrightarrow B$, i.e. mirror reflecting the figure along line $\sigma_{\rm A}=\sigma_{\rm B}$.  
		
\begin{figure}[h]
\begin{center}
\includegraphics[width=0.8\textwidth]{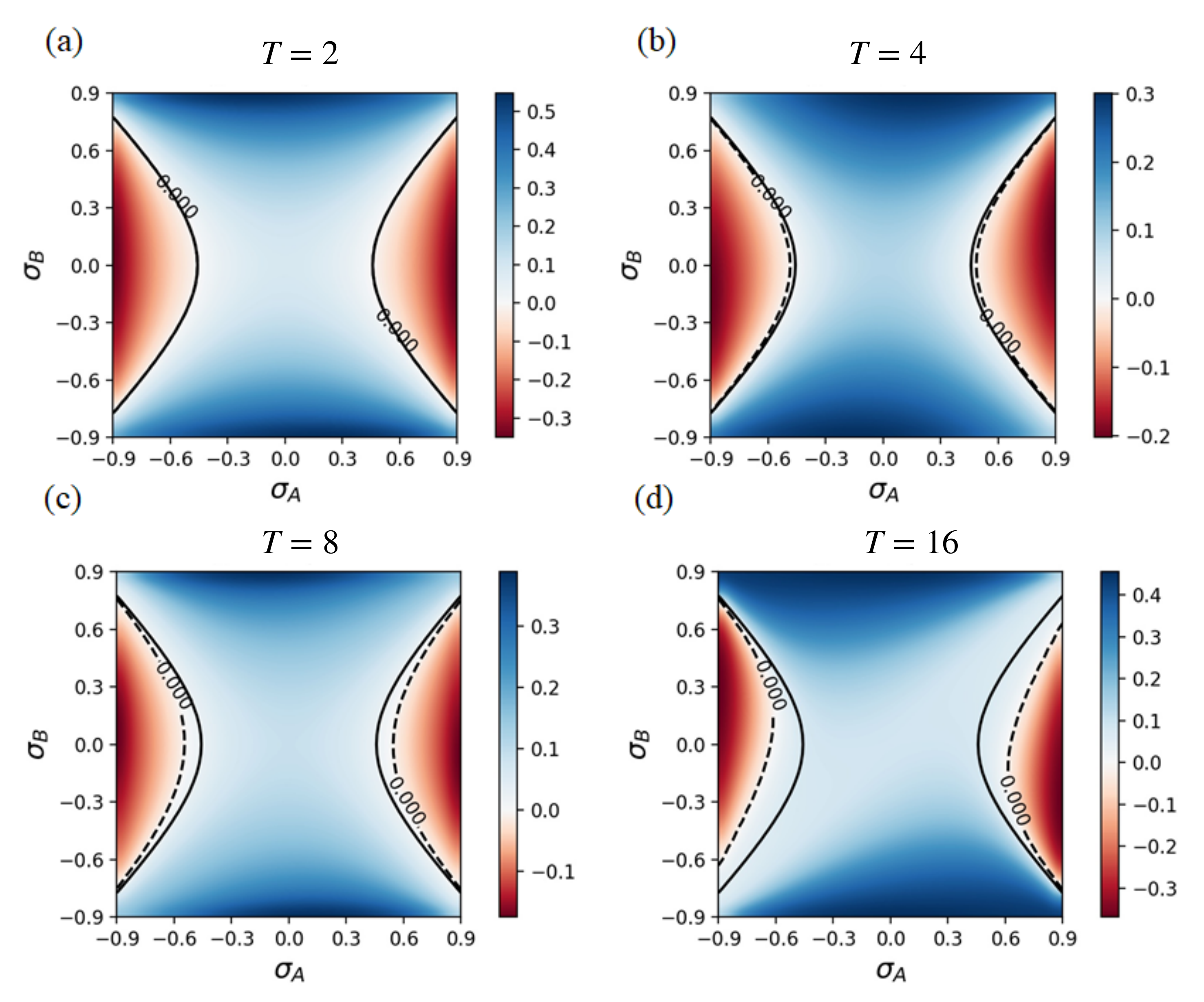}		
\caption{\textbf{Condition for selection of non-neutral mutants as fitness standard deviations, $\boldsymbol{\sigma_{\rm A}, \sigma_{\rm B}}$ vary.} The difference between the fixation probabilities, $\Delta \rho = \rho_A-\rho_B$, is plotted as $\sigma_A$ and $\sigma_B$ are varied for various periods $T$. (a)$T=2$, (b)$T=4$, (c)$T=8$ and (d)$T=16$. The size of the graph is $N=16$ and the inherent fitnesses are $r_A=1.1,\;r_B=1$. The dashed lines represent $\rho_A = \rho_B$ and the solid ones $\rho_A=\rho_B$ in a 2-chromatic cycle ($T=2$).}
\label{hm_diff_AB_1dot1_1_16}
\end{center}
\end{figure}

We have numerically solved the Kolmogorov equation for the fixation probabilities, $\rho_{\rm A}$ and $\rho_{\rm B}$, equations \ref{FR}-\ref{rho}, as a function of fitness standard deviations, $\sigma_{\rm A}, \sigma_{\rm B}$ for several values of $r_{\rm A}, r_{\rm B}$. Different sizes are considered and fixation probabilities for all $T$'s are computed. We compare our results with the exact result for $T=2$ case. The discrepancy between the curves of $\rho_A=\rho_B$ and Eq. \ref{condition_1}, becomes more significant as $T$ is increased. Figures \ref{hm_diff_AB_1dot1_1_16} and \ref{condition_SI_1} show the difference of the two fixation probabilities, $\Delta \rho = \rho_A-\rho_B$, for inherently advantageous and neutral mutants, respectively. The results are plotted for $N=16$ and different fitness periods $T=2,4,8,16$ in figure \ref{hm_diff_AB_1dot1_1_16}. 

For a uniform environment, i.e. the center point $(\sigma_{\rm A},\sigma_{\rm B})=(0,0)$ has $\rho_{\rm A} > \rho_{\rm B}$, since $r_{\rm A} = 1.1 > r_{\rm B} = 1$. As heterogeneity in mutant fitness, $\sigma_{\rm A}$, is increased at some point type A becomes disadvantaged. Figures \ref{hm_diff_AB_1dot1_1_16}(b-d) shows that for larger periods where the rich and poor sites are more segregated, the area of the region in $\sigma_{\rm A}-\sigma_{\rm B}$ plane where $\rho_{\rm A} < \rho_{\rm B}$, shrinks. This means that increasing  period, i.e. redistributing the fitnesses without changing mean or standard deviations, can make a deleterious mutant into an advantageous one (area between dashed and solid lines). However, this is not true for intrinsically neutral mutations $(r_{\rm A}=r_{\rm B}=1)$. As is seen from figure \ref{condition_SI_1} (see SI section), increase in period does not change the neutrality line away from the two diagonal lines $\sigma_{\rm A} = \pm \sigma_{\rm B}$. 

\begin{figure}[h]
\begin{center}
\includegraphics[width=0.9\textwidth]{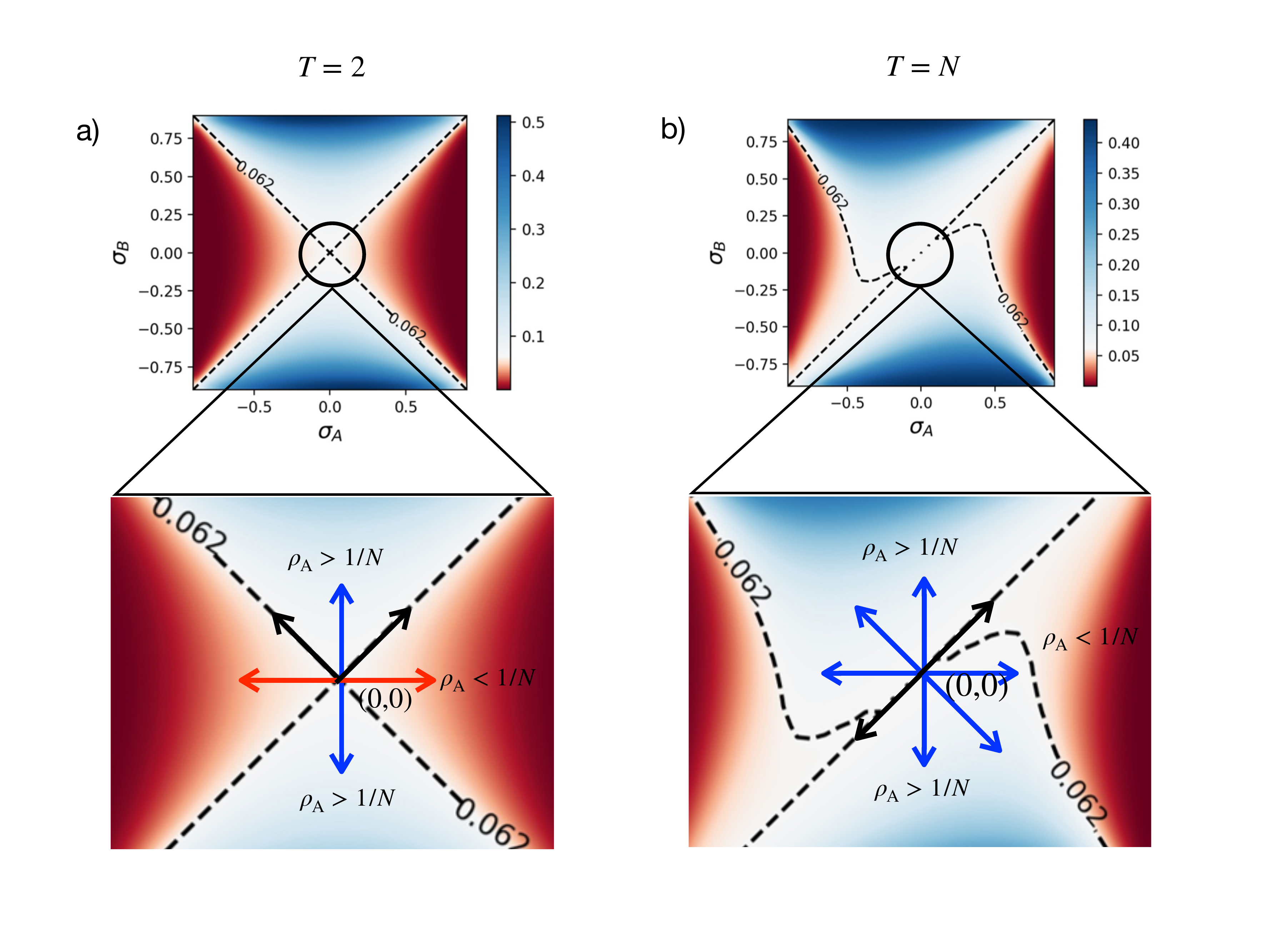}
\caption{\textbf{Fixation probability of a neutral mutant vs. standard deviations of fitnesses, $\boldsymbol{\sigma_A}$ and $\boldsymbol{\sigma_B}$.} Fixation probability of a mutant (type A) on a periodic cycle vs. $\sigma_A$ and $\sigma_B$ for small and large periods: (a)$T=2$, (b)$T=N$. The size of the graph is $N=16$ and the mutant is inherently neutral ($r_A=r_B=1$). The dashed lines represent the fixation probability of neutral mutant in the uniform environment ($1/N$). 
In bottom panels of (a) and (b) the blue arrows indicates the directions where $\rho_{\rm A}$ increases, red arrows shows the directions for which $\rho_{\rm A}$ decreases and black arrows indicates directions where $\rho_{\rm A}=1/N$ does not change. For large period $T=N$ almost introduction of any form of (weak) heterogeneity to the uniform model $(\sigma_{\rm A},\sigma_{\rm B})=(0,0)$ leads to increase in the fixation probability, $\rho_{\rm A}$ other than the symmetric interaction case (additive background fitness).}
\label{hm_prob_A_1_1_16}
\end{center}
\end{figure}

\begin{figure}[h]
\begin{center}
\includegraphics[width=0.9\textwidth]{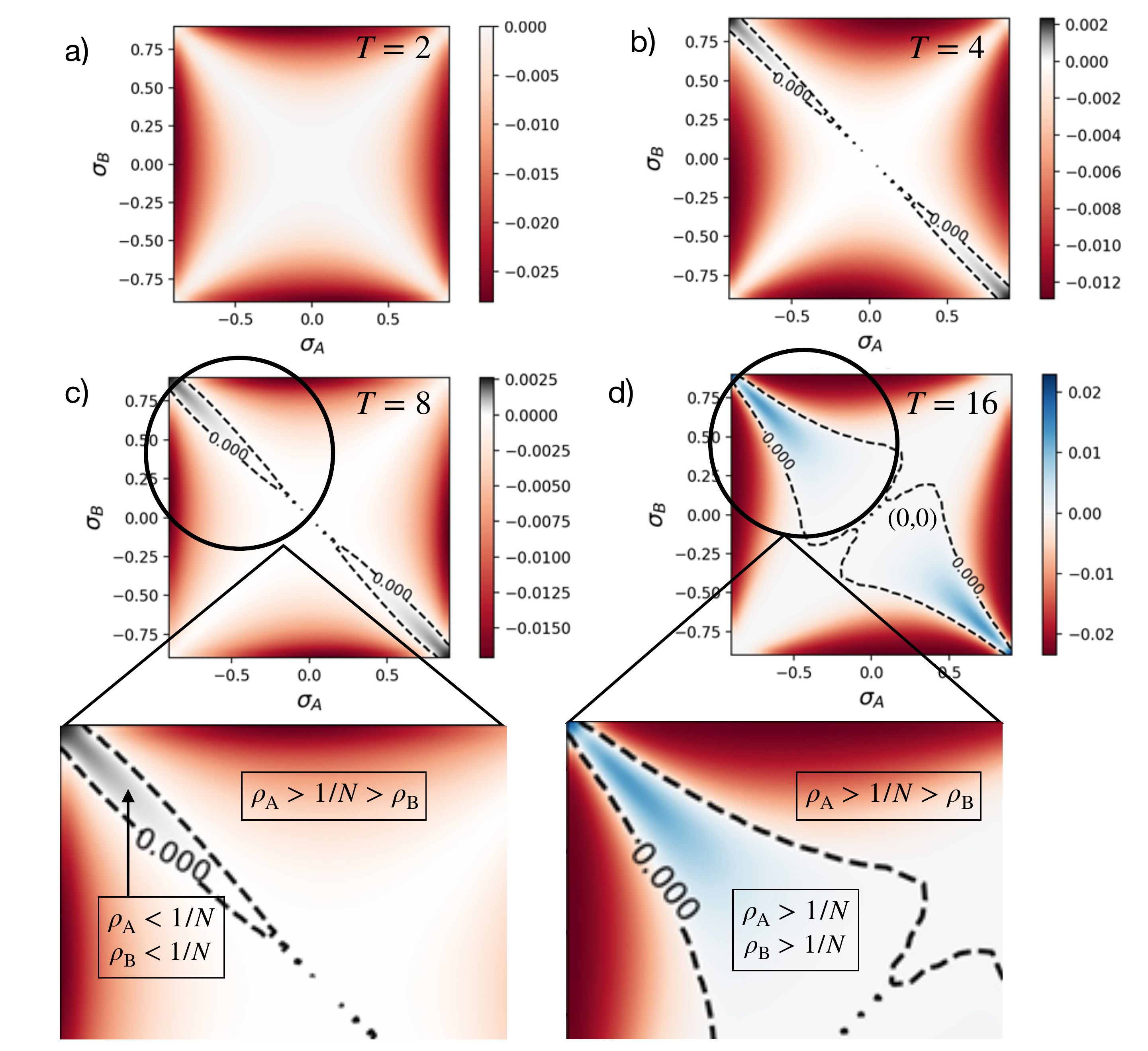}
\caption{$\boldsymbol{\zeta}$\textbf{ vs. $\boldsymbol{\sigma_A}$ and $\boldsymbol{\sigma_B}$.} The quantity $\zeta = (\rho_A-1/N)(\rho_B-1/N)$ on a periodic cyclic graph vs. $\sigma_A$ and $\sigma_B$ plotted for small to large periods. The periods are (a)$T=2$, (b)$T=4$, (c)$T=8$ and (d)$T=16$. $N$ is considered 16 and $r_A=r_B=1$. The sign of $\zeta$ determines the position of $\rho_{\rm A}$ and $\rho_{\rm B}$ compared to $1/N$. If $\zeta>0$, $\rho_{\rm A}$ and $\rho_{\rm B}$ are either greater or smaller than $1/N$, simultaneously. The blue color represents the area in which $\rho_A,\rho_B>1/N$ and the grey color represents $\rho_A,\rho_B<1/N$. On the dashed lines, at least one of $\rho_{\rm A}$ and $\rho_{\rm B}$ equals $1/N$. Finally, the red area where $\zeta<0$, shows the distributions in which one of $\rho_{\rm A}$ and $\rho_{\rm B}$ is greater and the other one less than $1/N$.}
\label{hm_mul_AB_1_1_16}
\end{center}
\end{figure}
	
We are also interested to see how changes in heterogeneity $\sigma_{\rm A}$ and/or $\sigma_{\rm B}$ 
affect
the chance of success of a mutant  in a heterogeneous environment compared to the case in a uniform environment. For brevity we focus on the neutral case $r_{\rm A} = r_{\rm B} = 1$ and ask how introducing arbitrary values of heterogeneity levels, $\sigma_{\rm A}$ and $\sigma_{\rm B}$ changes the fixation probability away from its uniform environment value, $\rho_{\rm A} = 1/N$. 

Figure \ref{hm_prob_A_1_1_16} shows the fixation probability of an inherently neutral mutant ($r_A=r_B=1$) in $\sigma_{\rm A}-\sigma_{\rm B}$ plane  for the smallest and the largest periods of fitness distribution ($T$). Panels a and b are plotted for periods $T=\;$2 and 16, respectively. The dashed lines represent the condition in which the fixation probability equals to the value pertaining to neutral mutation in the uniform environment i.e. $\rho_{\rm A}=1/N$. For $T=2$ (2-chromatic cycle) the above $\rho_{\rm A} = 1/N$ lines coincide with the lines $\sigma_A = \pm \sigma_B$. This can be readily observed by substituting $r_{\rm A}=r_{\rm B}$ in equation \ref{condition_1}. It is worth mentioning that fixation probability of type B, $\rho_{\rm B}$, can be similarly read off from figure \ref{hm_prob_A_1_1_16}. To do so, one just have to change the axes $\sigma_{\rm A}$ and $\sigma_{\rm B}$, due to the symmetry of $r_A=r_B=1$. 
	
As period is increased from minimum, $T=2$,  to medium values $T=4,8$, the $\rho_{\rm A}=1/N$ lines are not much changed from the two diagonal lines $\sigma_{\rm A} = \pm \sigma_{\rm B}$ (figure \ref{hm_prob_A_1_1_16}a). However, at maximal period $T=N$ the $\rho_{\rm A}=1/N$ line is drastically different from the other cases (figure \ref{hm_prob_A_1_1_16}b). This represents a case when environmentally different habitats are furthest from each other. In this case, we can see how introduction of small arbitrary heterogeneity in environmental interactions affects the condition for selection. In the neighborhood of $(\sigma_{\rm A},\sigma_{\rm B})=(0,0)$ (uniform environment) and for $T=N$, the fixation probability is always greater than $1/N$, except in a small window below diagonal line ($\sigma_A / \sigma_B=1- \epsilon~(0 \leq \epsilon \ll 1)$) that $\rho<1/N$. 
We have repeated this for larger $N$ and the same observation is hold true but the region with $\rho_{\rm A} <1/N$ shrinks even more. Meanwhile along the diagonal line $\sigma_{\rm A} = \sigma_{\rm B} = \sigma$, $\rho_{\rm A} = 1/N$ remains unchanged. Thus we conclude that most often a weak heterogeneity leads to increase in the fixation probability of a inherently neutral mutant. 

For strong heterogeneities we observe that there is an asymmetry between effect of heterogeneity on mutant and resident strategies. From figure \ref{hm_prob_A_1_1_16}a, we observe that along the horizontal line $\sigma_{\rm A}=\sigma, \sigma_{\rm B}=0$ (mutant fitness heterogeneity) the fixation probability is decreased, $\rho_{\rm A} < 1/N$. Conversely,  along the vertical line: $\sigma_{\rm A}=0, \sigma_{\rm B}=\sigma$ (resident fitness heterogeneity) the fixation probability is increased, $\rho_{\rm A} < 1/N$. This observation is consistent with similar observation on complete graph with fitness distribution \citep{kaveh2019environmental} or 2-chromatic graphs \citep{kaveh2020moran}. Similar results are also obtained for the case of inherently advantageous mutant ($r_{\rm A} = 1.1,r_{\rm B} = 1$), and inherently deleterious mutant ($r_{\rm A} =1, r_{\rm B}=1.1$). (See figures \ref{condition_SI_2} and \ref{condition_SI_3}.) 
			
We can also check the validity of the inequality $\rho_{\rm A} > 1/N > \rho_{\rm B}$ in a heterogeneous environment. In other words, we are interested to see if there are regions of plane $\sigma_{\rm A}-\sigma_{\rm B}$ plane for which we can have $\rho_{\rm A}$ and $\rho_{\rm B}$ simultaneously larger (or smaller) than $1/N$. To do so, we define the quantity $\zeta = (\rho_{\rm A} - 1/N) (\rho_{\rm B} - 1/N)$ as a function of $\sigma_{\rm A}$ and $\sigma_{\rm B}$. $\zeta$ is positive when both $\rho_{\rm A,B} < 1/N$ or both $\rho_{\rm A,B} > 1/N$ at the same time. Ref. \citep{kaveh2020moran} has shown that this does not happen for two-chromatic graphs ($T=2$ in our case). This is confirmed from figure \ref{hm_mul_AB_1_1_16}a. However as period increases, some regions in the $\sigma$-plane appear that have $\zeta > 0$. For medium-$T$ values, $T=4,8 (N=16)$ this is around the diagonal line $\sigma_{\rm A} = - \sigma_{\rm B}$. At maximum period $T=N$, the $\zeta > 1$ area covers a large range of $\sigma$-plane. Most significantly, we observe that around the uniform environment point $(\sigma_{\rm A} = 0, \sigma_{\rm B} = 0)$, we only have $\rho_{\rm A} > 1/N > \rho_{\rm B}$ (or $\rho_{\rm A} < 1/N < \rho_{\rm B}$) across the diagonal line $\sigma_{\rm A} = \sigma_{\rm B}$, corresponding to symmetric additive interactions. Otherwise, any other asymmetry in environmental interactions causes $\zeta > 0$. 

\begin{figure}
\begin{center}
\includegraphics[width=.8\textwidth]{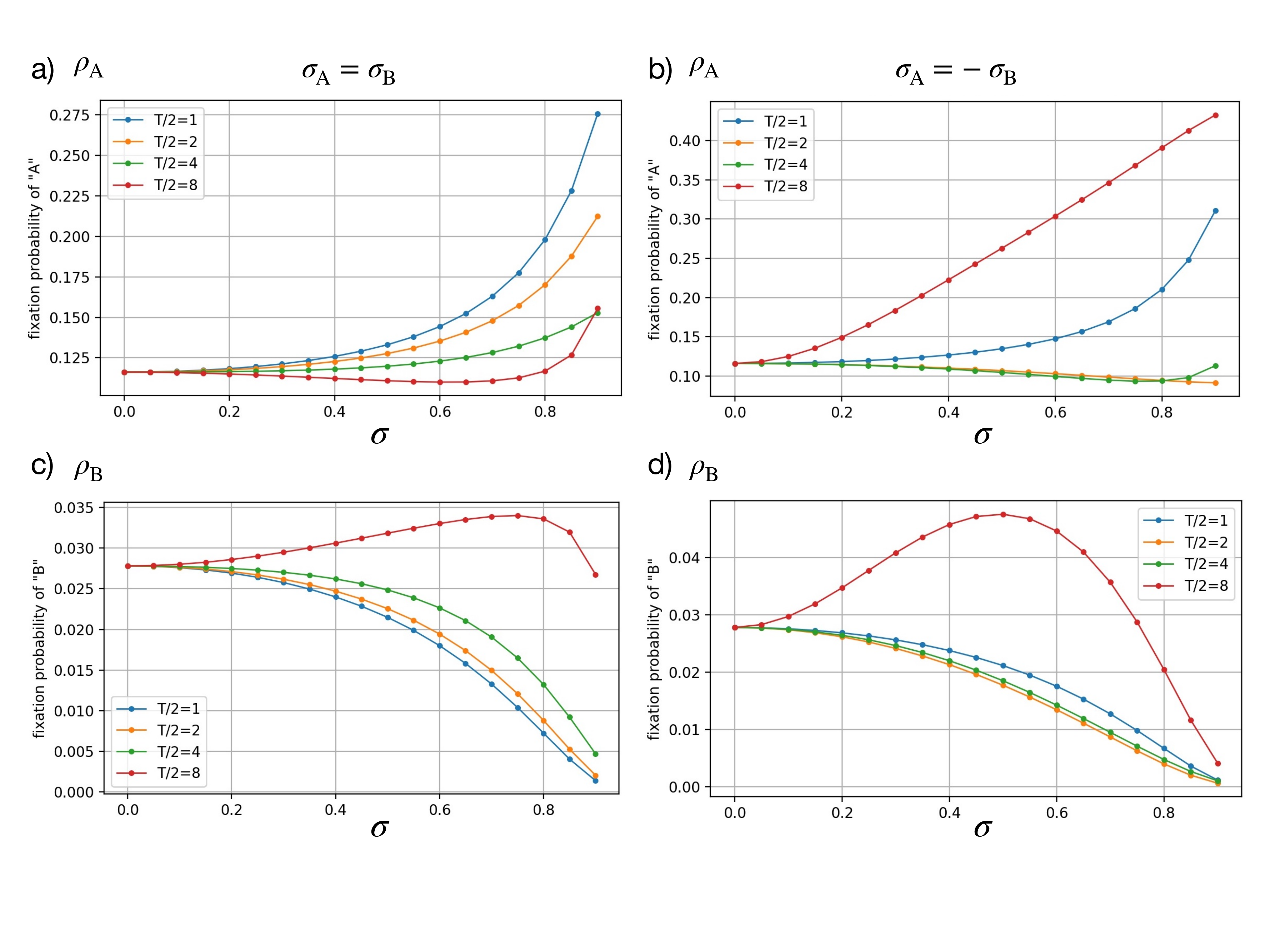}
\caption{\textbf{Fixation probability of an advantageous mutant vs standard deviation, $\sigma$, and period, $T$.} Fixation probability of an advantageous mutant (type A) plotted (a) vs. $\sigma$ for different periods for scenario 1 (\ref{scenario1}) and (b) Fixation probability of an advantageous mutant (type B) plotted (a) vs. $\sigma$ for different periods The graph size, $N$, is 16. In panels (c) and (d) the fixation probability of a type B in a background of type A's, $\rho_{\rm B}$, is plotted for both scenarios. The inherent fitnesses are $r_A=1.1,\;r_B=1$ and the fitness heterogeneity obeys the rule of Eq.\ref{scenario1} (a,c), Eq.\ref{scenario4}(b,d). The lines between the data points in both panels are plotted merely to guide the eye.}
\label{prob_A_1dot1_1_16}
\end{center}
\end{figure}

Finally, we plotted the fixation probabilities, $\rho_{\rm A}$ and $\rho_{\rm B}$ as a function of fitness standard deviation for two scenarios: completely symmetric environmental interaction scenario, equations \ref{scenario1} $\sigma_{\rm A}=\sigma_{\rm B}=\sigma$ and completely asymmetric interaction scenario, \ref{scenario4}, $\sigma_{\rm A}= -\sigma_{\rm B} = \sigma$. The two scenarios corresponds to the diagonal lines in the $\sigma_{\rm A}-\sigma_{\rm B}$ plane. The results are plotted for an inherently beneficial mutant, $r_{\rm A}=1.1, r_{\rm B}=1$, and for $N=16$ in figure \ref{prob_A_1dot1_1_16}. For scenario 1, it is observed that in general the fixation probability of type A increases as heterogeneity, $\sigma$ is increased. This is true for most periods, except $T=N$.  Behavior of $\rho_{\rm B}$ is opposite of that of $\rho_{\rm A}$. While $\rho_{\rm A}$ is increasing due to change in heterogeneity, $\rho_{\rm B}$ increases and vice versa. This is true for larger sizes $N=32, 64, 100$, as well. Conversely for completely asymmetric interaction scenario, scenario 2, the changes in $\rho_{\rm A}$ and $\rho_{\rm B}$ are in in tune. For some periods the fixation probability of both $\rho_{\rm A}$ and $\rho_{\rm B}$ can increase as $\sigma$ is increased (or vice versa). This is similar to previous observation made for the neutral case $r_{\rm A}=r_{\rm B}=1$, figure \ref{hm_mul_AB_1_1_16}b-d. 

	
\subsection{Time to fixation and coexistence}
\begin{figure}
\begin{center}
\includegraphics[width=0.8\textwidth]{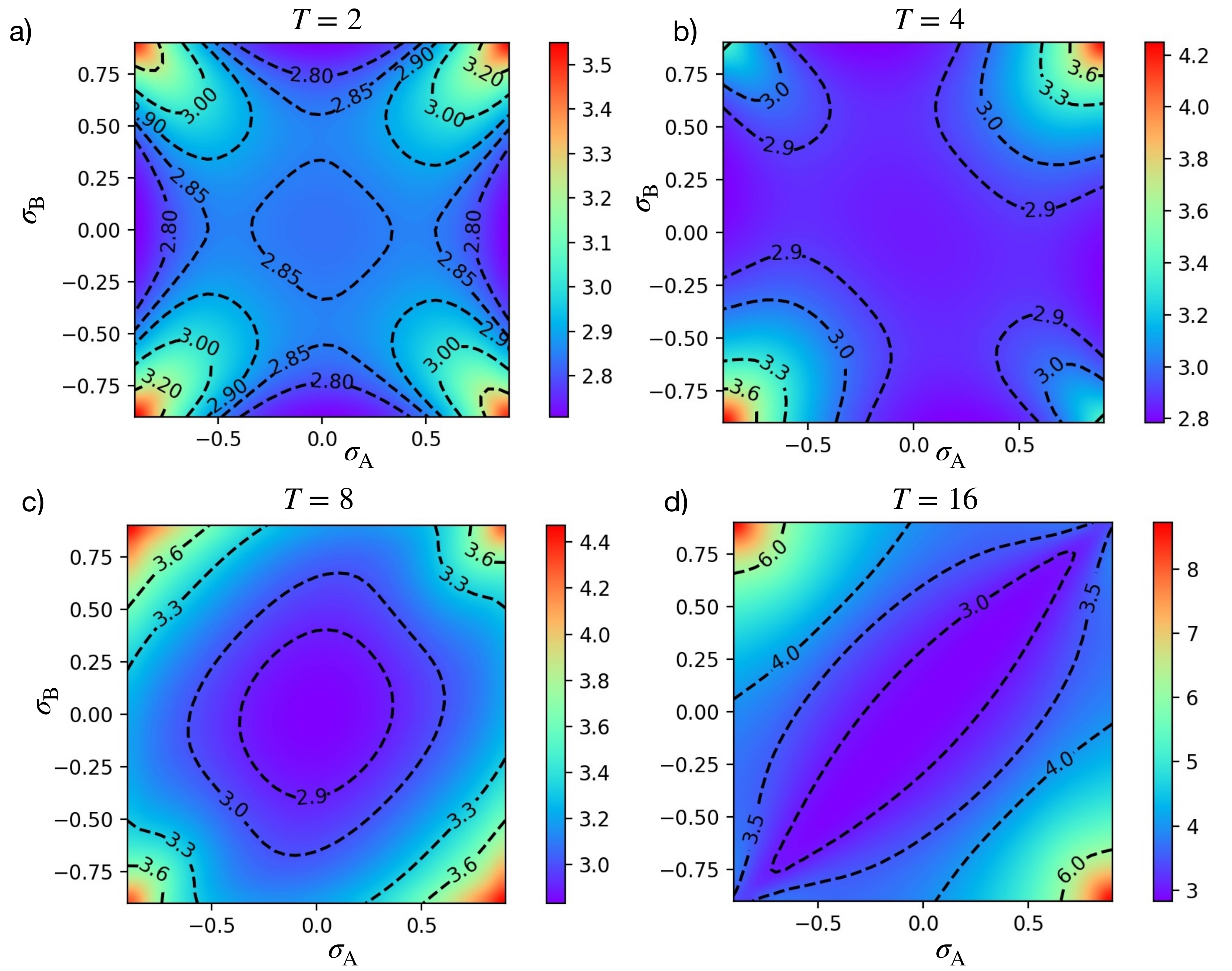}
\caption{\textbf{Fixation time as function of fitness standard deviations, $\sigma_{\rm A}, \sigma_{\rm B}$}. The fixation time for different periods is plotted as a function of $\sigma_{\rm A,B}$. Total population size is 16. $r_{\rm A} = r_{\rm B}=1$ and $N=16$. Notice for $T=N$ (panel d) time to fixation for $\sigma_{\rm A} \approx -\sigma_{\rm B} \approx 0.8$ increases $10^{6}$ fold relative to uniform environment fixation time, $(\sigma_{\rm A},\sigma_{\rm B})=(0,0)$.}
\label{fix_time_A_1dot1_1_16}
\end{center}
\end{figure}

We numerically calculated conditional mean time to fixation $\tau$. In figure \ref{fix_time_A_1dot1_1_16} we plotted $\log_{10}$ of average time to fixation in $\sigma_{\rm A}-\sigma_{\rm B}$ plane. Different panels indicate different periods. Equal $\tau$ contours plotted in the panels. For almost any of the periods, and along any direction in $\sigma$-plane the time to fixation generally increases with increase in heterogeneity (there are some exceptions, i.e., in the small periods and around the lines $\sigma_{\rm A}=0$ and $\sigma_{\rm B}=0$). However as the period is increased, time to fixation is increased much significantly along the $\sigma_{\rm A}=-\sigma_{\rm B}=\sigma$ axes which represents completely asymmetric interactions. For example at $\sigma_{\rm A}=-\sigma_{\rm B}\approx .8$ in the largest period, we have the $\tau \sim 10^{9}$ while for a uniform environment $\tau \sim 10^{3}$. This is 6 order of magnitude increase in time to fixation due to heterogeneity. We have repeated this for larger sizes and the ratio $\tau(\sigma=0.8)/\tau(\sigma=0)$ is much higher for $N=24, 32$.

\begin{figure}
\begin{center}
\includegraphics[width=0.8\textwidth]{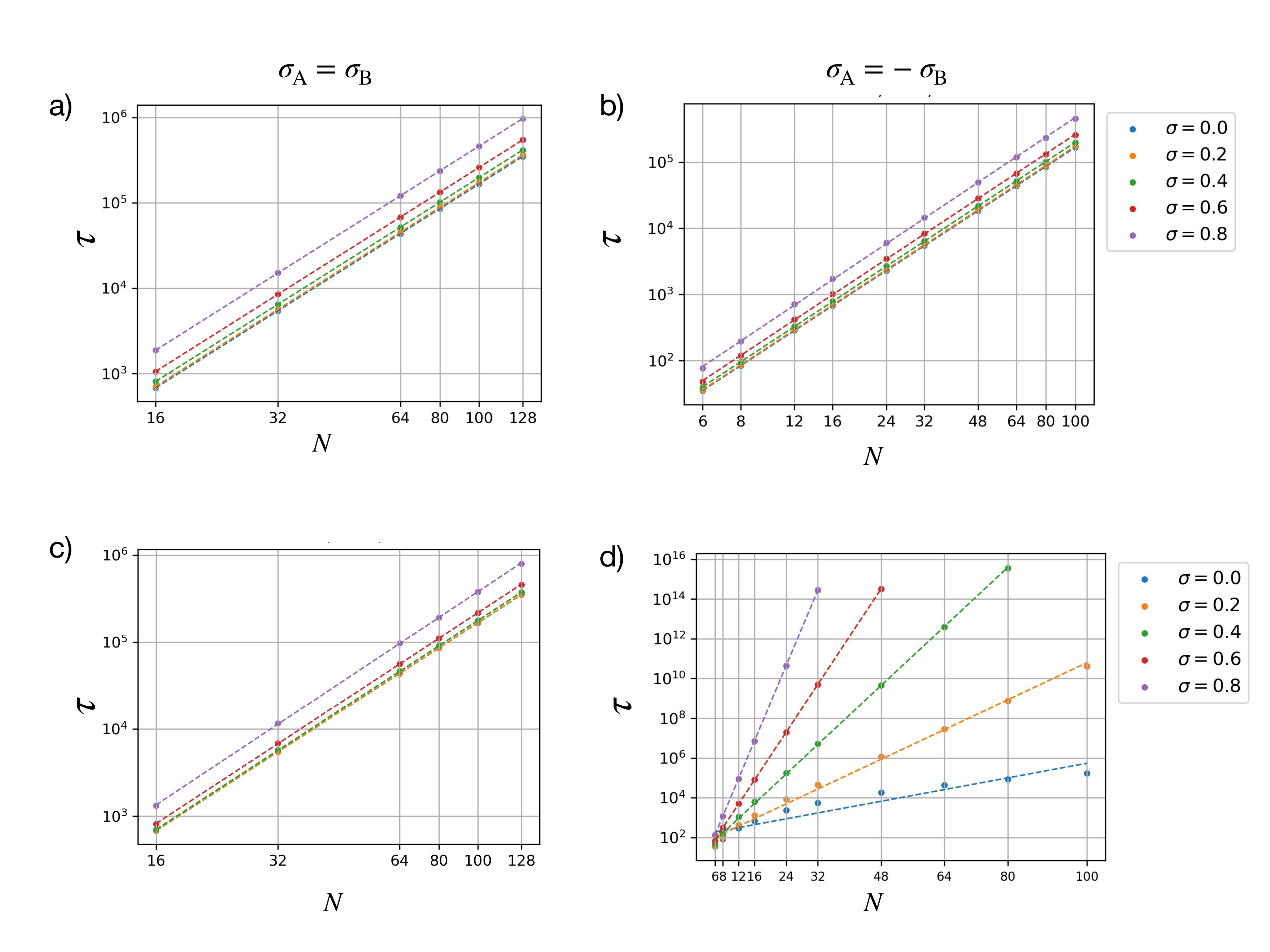}
\caption{\textbf{Scaling of fixation time with $\boldsymbol{N}$.} Fixation time is plotted vs. $N$ for the period $T=2,N$ in the case of $r_A=r_B=1$ and scenarios 1 ($\sigma_{\rm A}=\sigma_{\rm B}$) and 2 $(\sigma_{\rm A}=-\sigma_{\rm B})$ for some selected values of $\sigma=0,.2,.4,.6,.8$. a) Scenario 1 with minimum period $(T=2)$ (log-log plot). b) Scenario 2 with minimum period $(T=2)$ (log-log plot). c) Scenario 1 with maximum period $(T=N)$ (log-log plot). d) Scenario 2 with maximum period $(T=N)$ (semi-log plot). For (a)-(c) we observe power-law scaling $\tau \sim N^{\alpha}$ and for (d) we observe $\tau \sim \exp(\alpha N)$ at large $\sigma$'s.}
\label{scaling_1_1}
\end{center}
\end{figure}

For $r_{\rm A} = r_{\rm B}=1$ and symmetric interactions, scenario 1, there is a scaling relationship between population size and time to fixation, i.e. $\tau \sim N^{\alpha}$. Figure \ref{scaling_1_1} shows this fact. In the panel a and c, the fixation time $\tau$ in the symmetric interaction is plotted vs. the graph size $N$ for the periods $T=2$ and $T=N$, respectively. Some different values of $\sigma$ are being selected and the results are plotted in the log-log scale for the case of an inherently neutral mutant. The exponent of scaling for these two periods is between 3 and 3.1 and effectively equals to 3 for the uniform environment ($\sigma=0$).

For scenario 2, completely asymmetric environmental interactions, the scaling for the period $T=2$ as figure \ref{scaling_1_1}b shows, is in the form of  $\tau \sim N^{\alpha}$ with $\alpha\approx 3$, just like the symmetric scenario. However, this scaling does not hold for the largest period, i.e., $T=N$. For the largest period we can fit $\log_{10} \tau$ with $N$ itself. We already know that for $\sigma=0$ which defines a uniform environment, $\tau \sim N^{\alpha}$. However, as figure \ref{scaling_1_1}d shows, for the largest period in the fully asymmetric scenario, as $\sigma$ increases, the form of scaling gets close to $\tau \sim \exp{(\alpha N)}$. In other words for large enough values of $\sigma$, we have $\tau \sim \exp{(\alpha N)}$. The exponent $\alpha$ is a function of heterogeneity. With increasing $\sigma$, the value of $\alpha$ increases and besides, the exponential fitting gets more precise. This exponential behavior indicates a much more significant increase in the time to fixation. This combined with the curious increase in $\rho_{\rm A}$ and $\rho_{\rm B}$ with $\sigma$ for asymmetric interactions, suggest the possibility of coexistence of two types in realistic time scales in the stochastic models of the natural selection.

\subsection{Fitness shift effect}
We are interested to see if all the moments of fitness distributions for each of type A and type B were same, i.e. $r_{\rm A}=r_{\rm B}, \sigma_{\rm A}=\sigma_{\rm B}$ and other higher moments, would a mere reshuffling of fitnesses across the population lead to fitness advantage or disadvantage of the mutant type? We can achieve this by assuming same square-wave fitness functions as before but now $a_{i}$ and $b_{i}$ has a phase difference as originally suggested in previous section (See figure \ref{phase_diff_pattern}.)
		
\begin{figure}
\begin{center}
\includegraphics[width=0.5\textwidth]{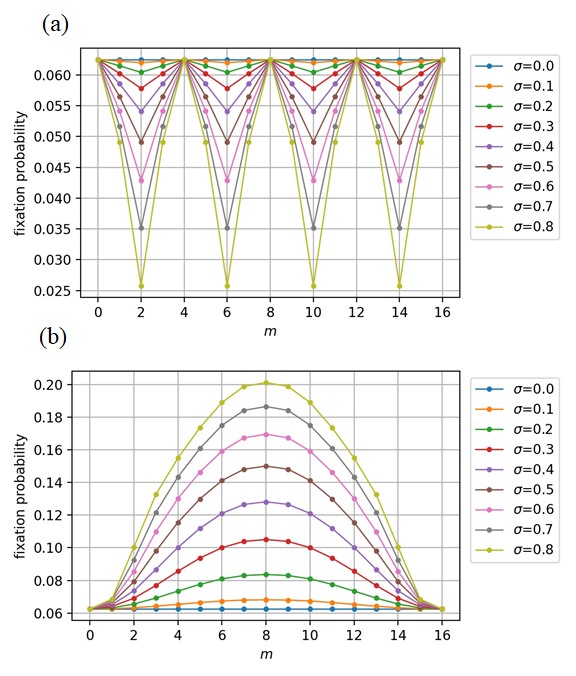}
\caption{\textbf{Effect of fitness shift on the fixation probability.} Fixation probability of an inherently neutral mutant is plotted vs. $m$ in the periods (a) $T=4$  and (b) $T=16$ for different values of $\sigma$. A 16-node cyclic graph is considered and at $m=0$, the rule of Eq. \ref{scenario1} is applied to the fitness pattern. The inherent fitnesses are considered to be $r_A=r_B=1$.}
\label{prob_shift}
\end{center}
\end{figure}

Figure \ref{prob_shift} depicts the fixation probability in two different periods of fitness distribution vs. the pattern shift $m$. The pattern shift, $m$, varies between 0 and $N$ ($N=16$). For the distribution period $T=4$ (panel a), one can see that a non-zero shift, decreases the fixation probability. Moreover, higher heterogeneity levels ($\sigma$) makes this reduction more significant. On the contrary, in the period $T=N$ (panel b) which is the largest possible length scale of fitness variations, adding a non-zero pattern shift, increases the fixation probability and the amount of this increment is more significant in higher $\sigma$'s. This means that depending on the length scale of the distribution, the pattern shift can either increase or decrease the fixation probability. The observations in figure \ref{prob_shift} are in agreement with figure \ref{hm_mul_AB_1_1_16}. If $m$ is an odd multiple of $T/2$, the heterogeneity pattern makes the environment be equivalent to the opposite fitness distribution ($\sigma_{\rm A}+\sigma_{\rm B}=0$) in figure \ref{hm_mul_AB_1_1_16}. One can see that on this line, for the middle distribution periods, the fixation probability of both types are smaller than uniform neutral case ($\rho_A,\rho_B<1/N$) and for the largest period greater than it ($\rho_A,\rho_B>1/N$). So, we have been expecting such behavioral difference between the two different coloring periods in figure \ref{prob_shift}. For the period $T=8$, the behavior is like the period $T=4$. For the smallest period, $T=2$, there is just one distinct state other than no-phase shift case, i.e., $m=1$ that this fitness shift is equivalent to opposite interaction distribution. For $T=2$, there is no difference between the fixation probability of the two possible states as they are both 2-chromatic graphs according to \citep{kaveh2020moran}. This is in agreement with what is shown in the figure \ref{hm_prob_A_1_1_16}a on the lines $\sigma_A-\sigma_B=0$ and $\sigma_A+\sigma_B=0$.

\begin{figure}
\begin{center}
\includegraphics[width=0.5\textwidth]{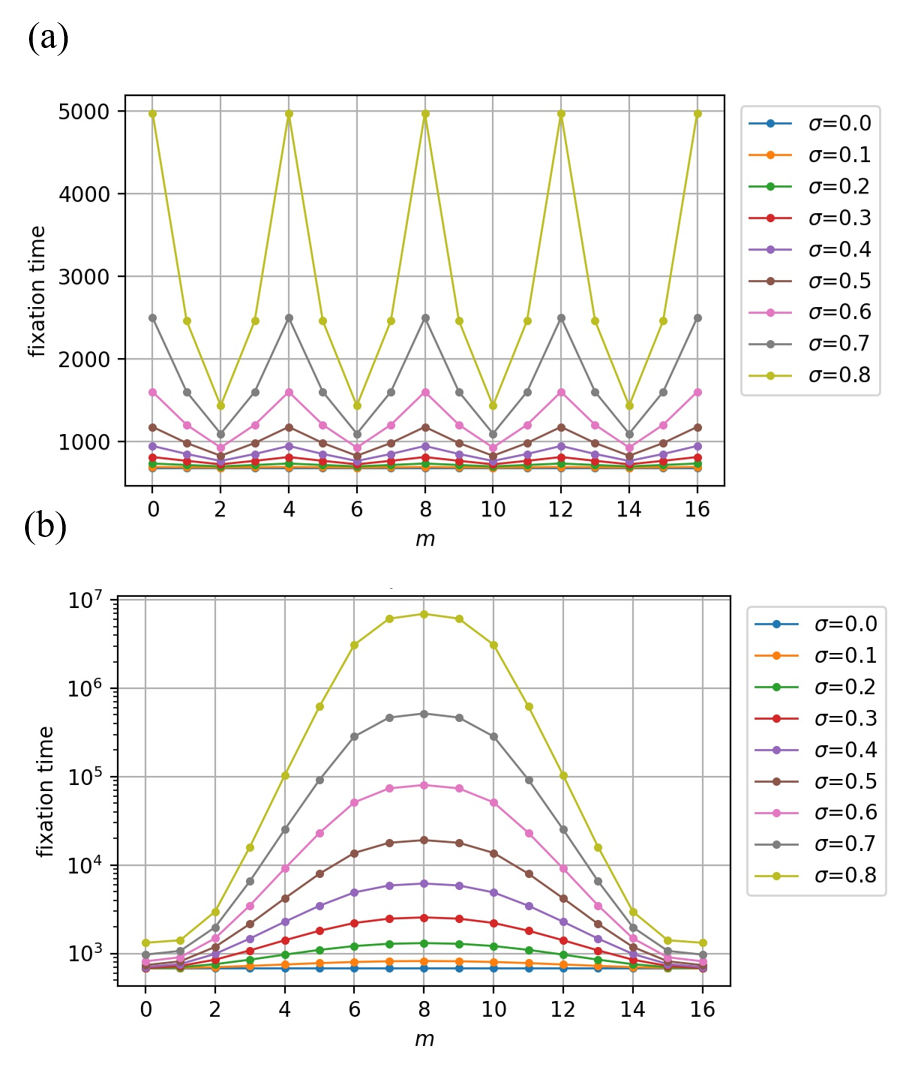}
\caption{\textbf{Effect of fitness shift on the fixation time.} Fixation time of an inherently neutral mutant is plotted vs. $m$ in the periods (a) $T=4$  and (b) $T=16$ for different values of $\sigma$. A 16-node cyclic graph is considered and at $m=0$, the rule of Eq. \ref{scenario1} is applied to the fitness pattern. The inherent fitnesses are considered to be $r_A=r_B=1$.}
\label{time_shift}
\end{center}
\end{figure}

In the same setup of figure \ref{prob_shift}, fixation time is studied when some pattern shift is imposed. In figure \ref{time_shift}, the fixation time of a 16-node cyclic graph is plotted vs. pattern shift, $m$, for two periods, panel a for $T=4$ and panel b for $T=16$. It can be seen that for $T=4$, the fixation time decreases for every equivalently non-zero value of $m$, while increases for $T=16$. Like the fixation probability, the direction of changing is dependent on the distribution period. However, the variations in all periods get more drastic as the heterogeneity amplitude increases and the phase shift gets close to $\pi/2$. Besides, from figure \ref{time_shift}b we learn that fixation time can increase several orders of magnitude in the case of $\pi/2$ phase shift (equivalent to opposite interaction fitness) compared to zero phase shift, if the heterogeneity period and amplitude are large enough.		
	
\section{Concluding remarks}

Evolutionary dynamics in network structures with spatial environmental heterogeneity has been of much interest in recent literature. Variations in environmental interactions can be due to uneven distribution of a resource, such as a nutrient, across the population. Environmental interactions can be symmetric or asymmetric relative to each genotype. Environmental interaction introduces an additive factor to the inherent fitness of each type. The question is how heterogeneity in fitness distribution influences natural selection in finite populations. Heterogeneity is modeled with standard deviation of the fitnesses, $\sigma_{\rm A}, \sigma_{\rm B}$ as well the spatial distribution of fitnesses. 

In this work we studied natural selection assuming general environmental interactions in linear spatial structures and periodic spatial fitness distributions. Periodic fitness distributions cover two important biological limits. Maximum periods corresponds to the cases where one side of the population is under one environmental condition and the other side is another environmental condition. This is a two-habitat subdivided population model where inside each habitat we keep the underlying graph population structure. Minimum period, can be used as an approximation of fine-grained random fitness heterogeneities due to imperfect resource distributions.   

We also considered the condition for selection in this setting. This is defined by $\rho_{\rm A} = \rho_{\rm B}$. If the mutant is inherently beneficial, the condition for selection is $\sigma_{\rm A} < \sigma_{\rm B}$. It does not change as period is varied. However, for inherently beneficial or deleterious mutants, this chages. Increase in the period can change some deleterious mutants into beneficial ones. 

We calculated the fixation probability of a random mutant $\rho_{\rm A}$ and average time to fixation, $\tau$, as functions of $\sigma_{\rm A}$, $\sigma _{\rm B}$ as well as period $T$. For inherently neutral mutants, $r_{\rm A} = r_{\rm B}$, the fixation probability in the uniform environment $\rho_{\rm A} = 1/N$ either decreases (increases) if $\sigma_{\rm A} > \sigma_{\rm B}~(\sigma_{\rm A} < \sigma_{\rm B})$ respectively. For large periods, however, almost any form of heterogeneity lead to increase in the fixation probability. Similar observations are made for beneficial and deleterious mutants. For example, resident-specific environmental interactions, i.e. $\sigma_{\rm A}=0, \sigma \neq 0$ always lead to increase $\rho_{\rm A}$ while mutant-specific interaction always reduce $\rho_{\rm A}$. Interestingly, in some regions of the $\sigma_{\rm A}-\sigma_{\rm B}$ plane the fixation probability of both type A and type B is increased at the same time. In other words, we can have regions of the parameter space that $\rho_{\rm A} > 1/N$ and $\rho_{\rm B} > 1/N$. 

Average time to fixation, $\tau$, increases overall with heterogeneity in all periods with very few exceptions. Fixation time shows a scaling with the graph size in the neutral types case and symmetric interactions. 
For completely asymmetric interactions and large periods, we observe exponential scaling in the fixation probability. This exponential increase in time to fixation coincides with regimes where $\rho_{\rm A} > 1/N$ and $\rho_{\rm B} > 1/N$, and indicates potential for coexistence in the realistic time scales. 

If the fitness pattern of types A and B have some phase difference compared to symmetric interaction heterogeneity model, the fixation probability and fixation time are changed. The change gets more significant as the phase shift closes to $\pi/2$ and the heterogeneity amplitude $\sigma$ increases. For instance, in the largest period, every value of phase shift increases the fixation probability and fixation time compared to the case of no phase shift. In general, the direction of $\rho$ and $\tau$ variation due to phase shift depends on the period of distribution.
	
\section*{Acknowledgment}
	The authors are grateful to Jamal Doroud for technical support and HPC of Sharif University of Technology.


\bibliographystyle{spbasic}
\bibliography{citations}	

\clearpage	
	
\appendix
\section{Elements of matrices Q and R}
\label{Q_R_elemants}
	
\begin{figure}[h]
\includegraphics[width=0.4\textwidth]{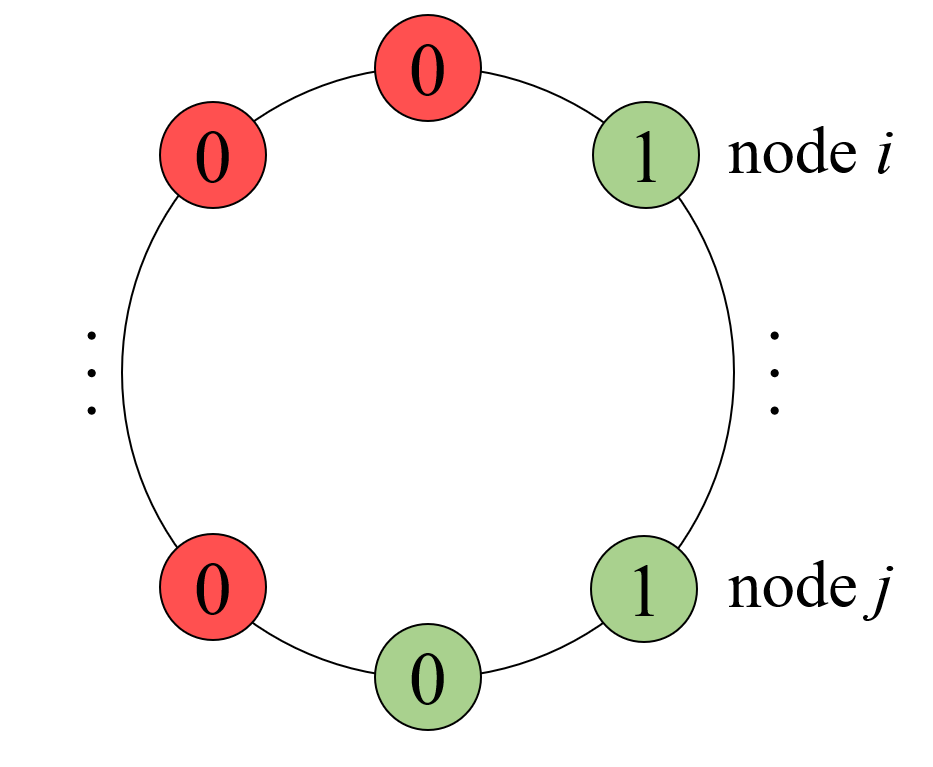}
\caption{{\bf Configuration of mutants and residents on a cycle.} Every configuration of mutants and residents on a cyclic graph can be determined by mentioning the beginning ($i$th) and the end ($j$th) node of the mutants chain, which we have to go from beginning to the end node in a specific direction (say, CW)  to cover the mutants. Green and red colors indicate rich and poor nodes (See text).}
\label{model_fig}
\end{figure}

	If we calculate the elements of the matrices $\bold{Q}$ and $\bold{R}$, which are the transition probabilities between different states, there will be need to just some matrix manipulations to solve the problem. As an example, we show how to calculate the transition probabilities between some states. We are studying cyclic graphs. On a cyclic graph, when a mutant is put in a random node and the dynamics of Moran process begins, always the mutants form a connected subset of the graph. Thus, we can completely determine every possible configuration by denoting just the beginning and the end nodes of the mutant chain, like what is shown in figure \ref{model_fig}. In this figure, the number written in each node, determines the number of mutants inside it. Nodes $i$ and $j$ are the begin and end nodes of the mutant chain. When referring every node label, we should be aware of periodicity. Thus, node 1 comes after node $N$.

	Let's suppose that the fitness of mutant and resident types in every node $k$ ($k \in \{1,...,N\}$) is $a_k$ and $b_k$, respectively. If there are more than one element of each type in the graph, the current and the next states are both transient and all the possible transitions are between two transient states. Thus the probabilities of such transitions must be inserted into the matrix $\bold{Q}$. In such a situation that is shown in figure \ref{model_fig}, there is just 5 possible transitions in the next time step:
	\begin{enumerate}
		\item $n_{j+1}=0 \to n_{j+1}=1$:
		\begin{equation*}
		P_1= {a_j/2 \over \sum_{k=1}^{N}{(a_k n_k +b_k (1-n_k))}}
		\end{equation*}
		\item $n_{j}=1 \to n_{j}=0$:
		\begin{equation*}
		P_2= {b_{j+1}/2 \over \sum_{k=1}^{N}{(a_k n_k +b_k (1-n_k))}}
		\end{equation*}
		\item $n_{i-1}=0 \to n_{i-1}=1$:
		\begin{equation*}
		P_3= {a_{i}/2 \over \sum_{k=1}^{N}{(a_k n_k +b_k (1-n_k))}}
		\end{equation*}
		\item $n_{i}=1 \to n_{i}=0$:
		\begin{equation*}
		P_4= {b_{i-1}/2 \over \sum_{k=1}^{N}{(a_k n_k +b_k (1-n_k))}}
		\end{equation*}
		\item nothing changes (the birth and death happens inside the interior zone of either two types):
		\begin{equation*}
		P_5= 1-\sum_{l=1}^{4}{P_l}
		\end{equation*}
	\end{enumerate}
	The transition probability to the other configurations is zero.
	
	On the other hand, if any of the next time step probable states is either fully resident or fully mutant state, the pertinent transition probability must be inserted into the matrix $\bold{R}$.
	
	So, using this simple approach, we can find the transition probability between any two states and consequently, the matrix $\bold{T}$ can be completely made up.
	
\section{Supplementary figures and tables}

\begin{figure*}
\includegraphics[width=0.7\textwidth]{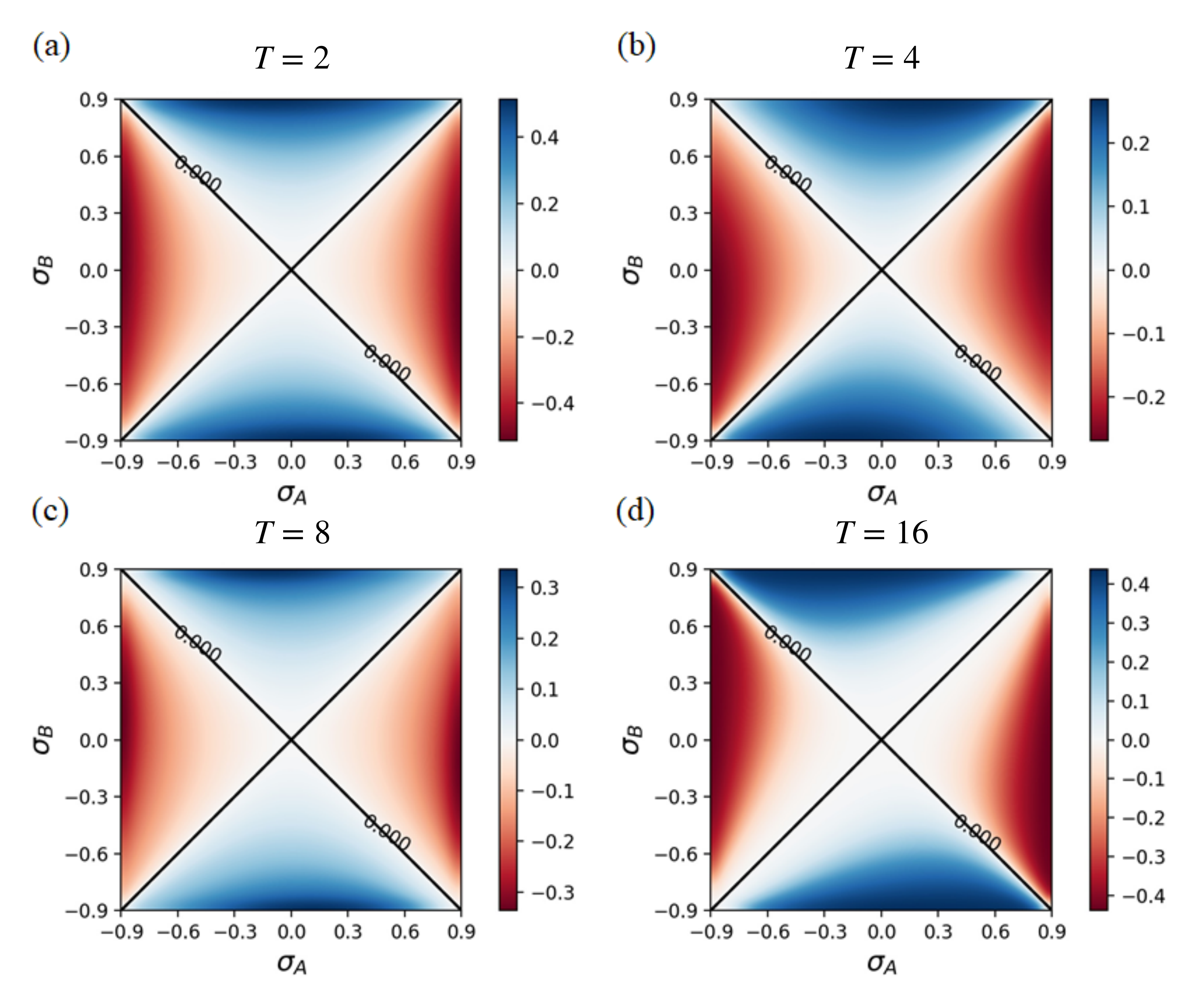}
\caption{$\boldsymbol{\rho_A-\rho_B}$ \textbf{of neutral mutants vs. $\boldsymbol{\sigma_A}$ and $\boldsymbol{\sigma_B}$.} $\rho_A-\rho_B$ on a periodic cycle vs. $\sigma_A$ and $\sigma_B$ for small to large periods ($T$ denotes the period): (a)$T=2$, (b)$T=4$, (c)$T=8$ and (d)$T=16$. The size of the graph is $N=16$ and the mutant is inherently neutral ($r_A=r_B=1$). The dashed lines represent $\rho_A-\rho_B=0$ and the solid ones $\rho_A-\rho_B=0$ for the case of 2-chromatic graph ($T=2$).}		
\label{condition_SI_1}
\end{figure*}
	
\begin{figure*}
\includegraphics[width=0.7\textwidth]{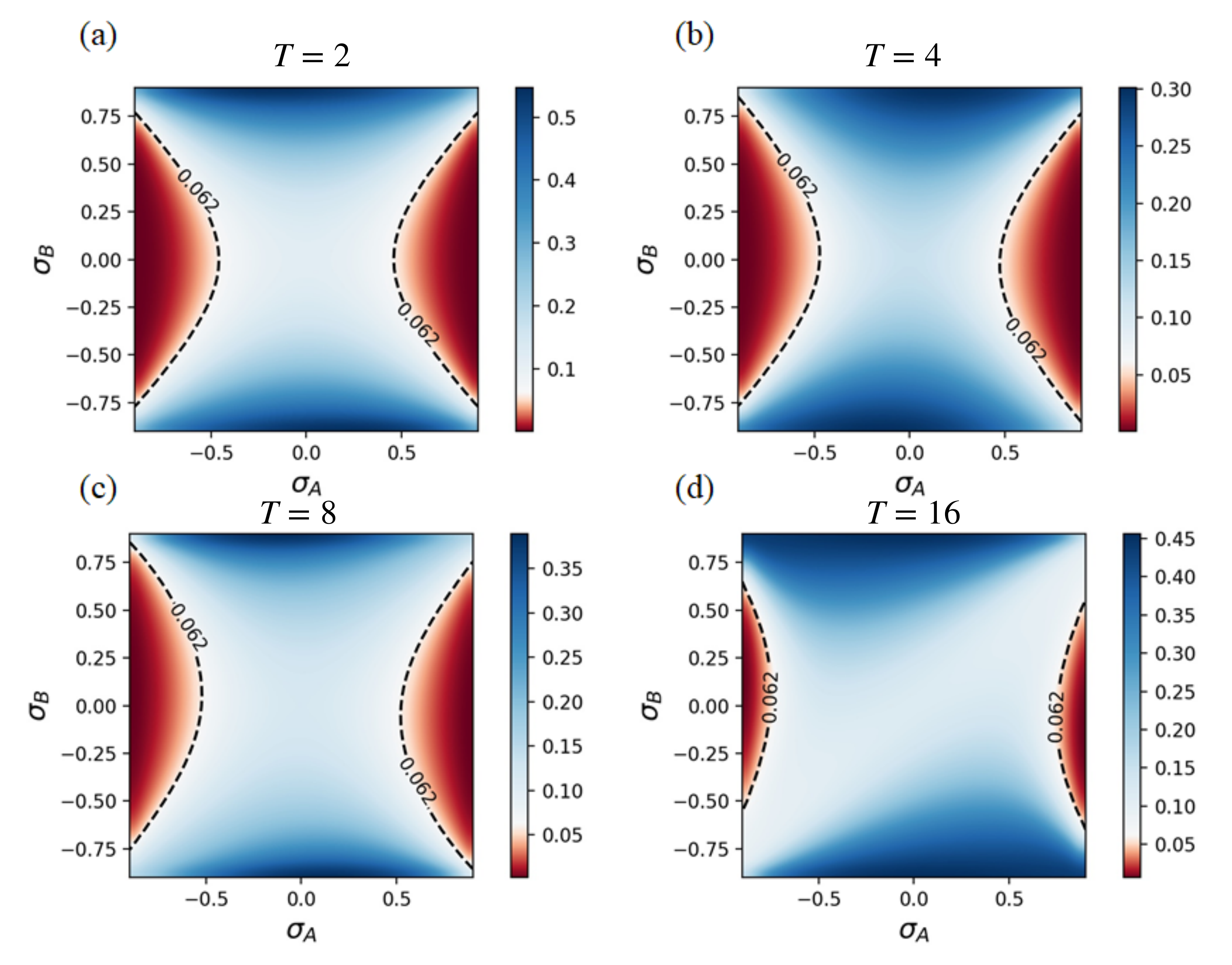}
\caption{\textbf{Fixation probability of an advantageous mutant vs. $\boldsymbol{\sigma_A}$ and $\boldsymbol{\sigma_B}$.} Fixation probability of a mutant (type A) on a periodic cycle vs. $\sigma_A$ and $\sigma_B$ for small to large periods: (a)$T=2$, (b)$T=4$, (c)$T=8$ and (d)$T=16$. The size of the graph is $N=16$ and the mutant is inherently neutral ($r_A=1.1,\;r_B=1$). The dashed lines represent the fixation probability of neutral mutant in the uniform environment ($1/N$).}
		\label{condition_SI_2}
	\end{figure*}
	
\begin{figure*}
\includegraphics[width=0.7\textwidth]{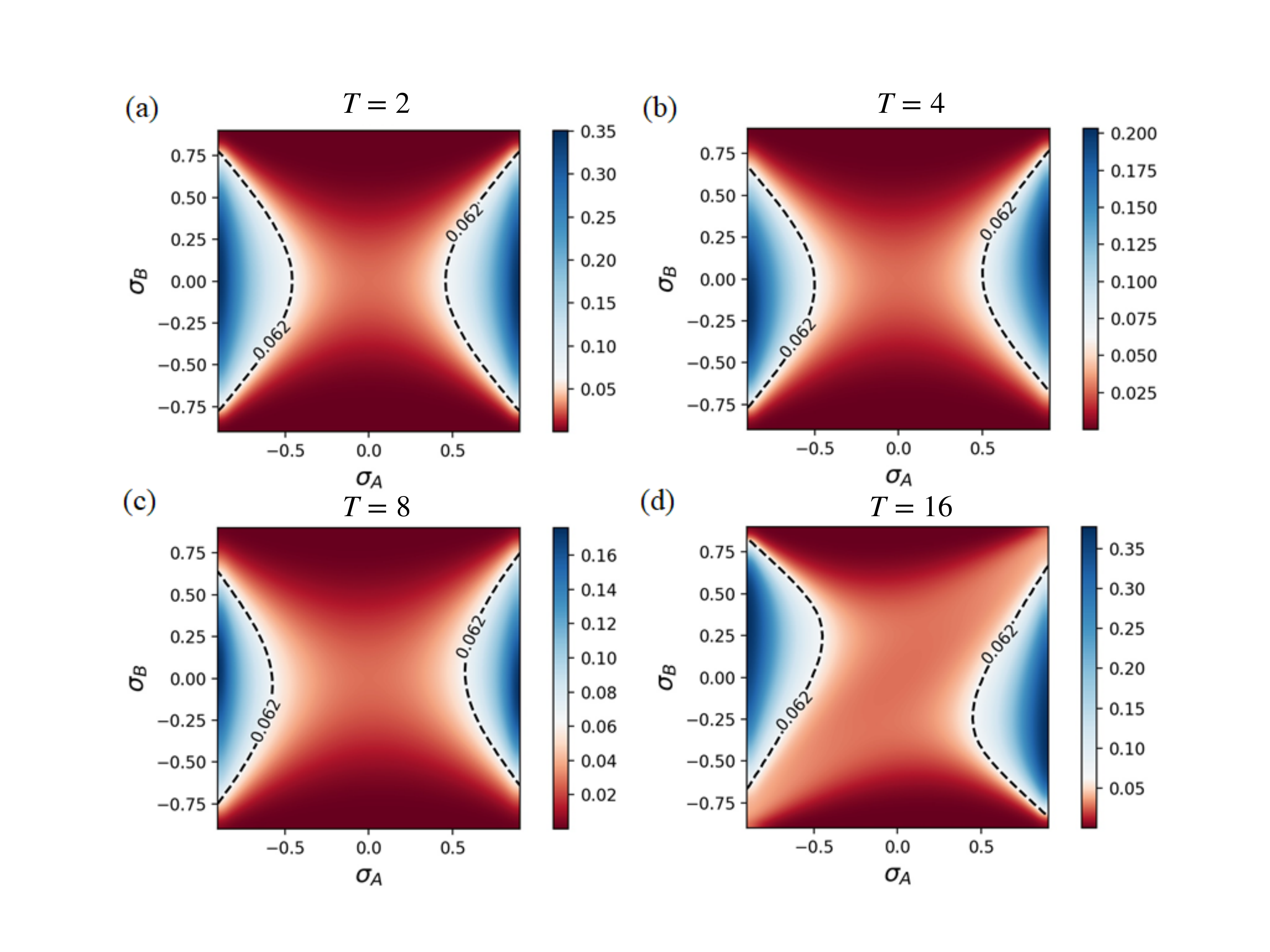}
\caption{\textbf{Fixation probability of a deleterious mutant vs. $\boldsymbol{\sigma_A}$ and $\boldsymbol{\sigma_B}$.} Fixation probability of a deleterious mutant on a periodic cycle vs. $\sigma_A$ and $\sigma_B$ for small to large periods ($T$ denotes the period): (a)$T=2$, (b)$T=4$, (c)$T=8$ and (d)$T=16$. The size of the graph is $N=16$ and the mutant is inherently deleterious ($r_A=1,\;r_B=1.1$). The dashed lines represent the fixation probability of neutral mutant in the uniform environment ($1/N$).}
\label{condition_SI_3}
\end{figure*}	
		
%


\end{document}